\documentclass{PoS}
\usepackage[percent]{overpic}
\usepackage{color}

\title{Zooming towards the Event Horizon - mm-VLBI today and tomorrow}

\ShortTitle{mm-VLBI today and tomorrow}

\author{\speaker{Thomas P. Krichbaum}, A. Roy, J. Wagner, H. Rottmann, J.A. Hodgson, A. Bertarini, W. Alef, J.A. Zensus \\
        Max-Planck-Institut f\"ur Radioastronomie, Auf dem H\"ugel 69, Bonn, Germany\\
        E-mail: \email{tkrichbaum@mpifr.de}}

\author{A.P. Marscher, S.G. Jorstad\\
        Institute for Astrophysical Research, Boston University, Boston, MA, USA}

\author{R.~Freund, D.~Marrone, P.~Strittmatter, L.~Ziurys\\
       Arizona Radio Observatory, Tucson, Arizona 85721, USA}

\author{R.~Blundell, J.~Weintroub, K.~Young\\
       Harvard Smithsonian Center for Astrophysics, Cambridge, MA 02138, USA}

\author{V.~Fish, S.~Doeleman\\
       MIT Haystack Observatory, Westford, MA 01886, USA }

\author{M. Bremer, S. Sanchez\\
       Institut de Radioastronomie Millim\'etrique (IRAM) France and Spain} 

\author{L. Fuhrmann, E. Angelakis, V. Karamanavis \\
       Max-Planck-Institut f\"ur Radioastronomie, Bonn, Germany}

\abstract {Global VLBI imaging at millimeter and sub-millimeter wavelength overcomes
the opacity barrier of synchrotron self-absorption in AGN and opens the direct view
into sub-pc scale regions not accessible before. Since AGN variability is more pronounced at short 
millimeter wavelength, mm-VLBI can reveal structural changes in very early stages after
outbursts. When combined with observations at longer wavelength, global 3\,mm and 1\,mm
VLBI adds very detailed information on the source structure. This helps 
to determine fundamental physical properties at the jet base, and in the vicinity 
of super-massive black holes at the center of AGN.
Here we present new results from multi-frequency mm-VLBI imaging of OJ\,287 during a major outburst.
We also report on a successful 1.3\,mm VLBI experiment with the APEX telescope in Chile.
This observation sets a new world record in angular resolution. It also opens
the path towards future mm-VLBI with ALMA, which aims at the mapping of the black hole event
horizon in nearby galaxies, and the study of the roots of jets in AGN.
}

\FullConference{11th European VLBI Network Symposium \& Users Meeting,\\
		October 9-12, 2012\\
		Bordeaux, France}

\begin{document}

\section{Introduction}
\vspace*{-0.4cm}
In Active Galactic Nuclei (AGN) the energy extraction from black holes (BH) and the 
detailed understanding of jet formation and jet acceleration are still poorly understood.
It is therefore desirable to observe galaxies with black holes 
and their emanating jets with an as high as possible angular and spatial resolution. 
For nearby super massive black holes, such as the BH in the Galactic Center 
(D $\sim 8$\,kpc) and in M\,87 (Virgo A, D $\sim 17$\,Mpc) ground-based VLBI 
at $\lambda \leq 1.3$\,mm ($\nu \geq 230$\,GHz) provides an observing beam 
well matched to the expected size of the event horizon - and by this to the size of 
the observable emission region around such black holes. For the more distant quasars
(and other AGN) a spatial resolution of several 10-1000 gravitational radii is obtained.
Two fundamentally different types of jet formation models are often discussed:
(i) in Blandford-Payne (BP) type models the particle acceleration is done via a magnetic
sling-shot mechanism with field lines anchored in the rotating accretion
disk collimating and accelerating a disk wind. (ii) In Blandford-Znajek (BZ) type models the
energy extraction is purely electromagnetic and is directly coupled to the spin 
of the black hole. As a consequence of this, the diameter and morphology of the jet base (i.e. at radii
where the jet emission becomes observable) may be different for BP- and BZ-type jet launching,
in the sense that BZ-jets may appear more narrow and compact. VLBI imaging with ten micro-arcsecond scale 
resolution can help to discriminate between these models and constrain the
parameter space for the theoretical modeling.


\section{Global VLBI at 3\,mm wavelength}
\vspace*{-0.4cm}
Global 3\,mm VLBI imaging at 86\,GHz is performed with the stand-alone VLBA (8 antennas
equipped with 3\,mm receivers, no receiver at HN and SC) and with the Global Millimeter VLBI
Array (\href{http://www.mpifr-bonn.mpg.de/div/vlbi/globalmm/}{GMVA}). The GMVA combines the
big European antennas (100\,m Effelsberg, 30\,m IRAM Pico Veleta, (6x15)\,m IRAM Plateau de Bure, 40\,m Yebes,
20\,m Onsala, 14\,m Mets\"ahovi) with the VLBA. While the VLBA
could observe more frequently, the GMVA is about 3 times more sensitive and offers also
a much higher resolution than the VLBA (up to 40\,$\mu$as). A brief description of the present
status of the GMVA and new results are presented by J. Hodgson et al., in this conference 
\cite{Hodgson} (and references therein).

\begin{figure}
\includegraphics[trim= 65mm 0mm 50mm 0mm, clip, width=0.3\textwidth]{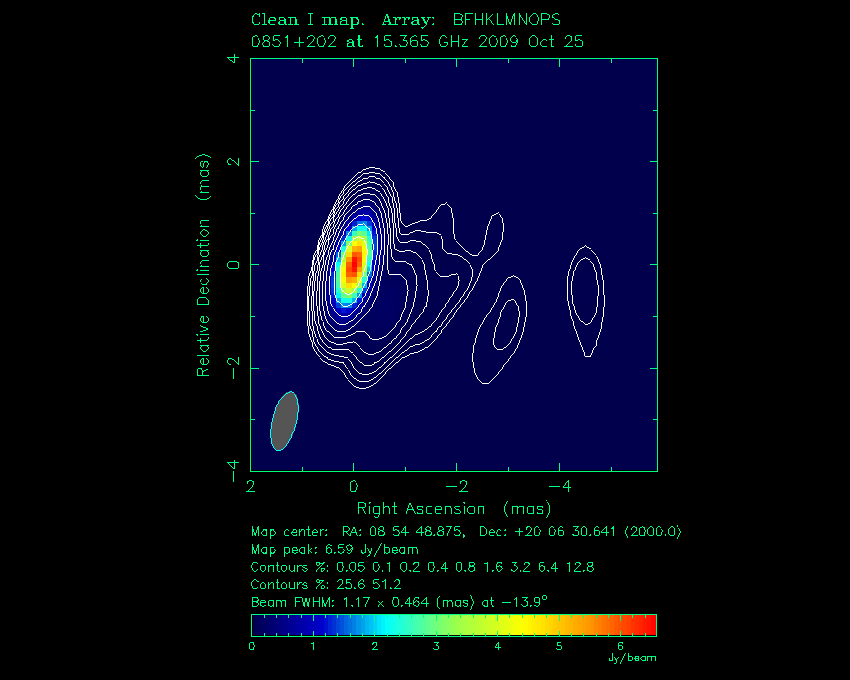}
\includegraphics[trim= 65mm 0mm 50mm 0mm, clip, width=0.3\textwidth]{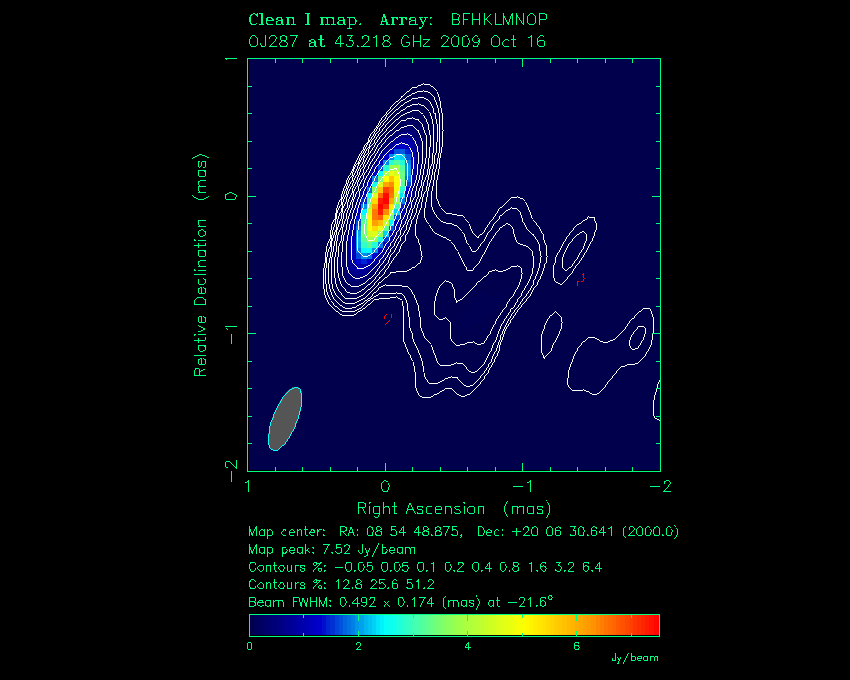}
\includegraphics[trim= 65mm 0mm 50mm 0mm, clip, width=0.3\textwidth]{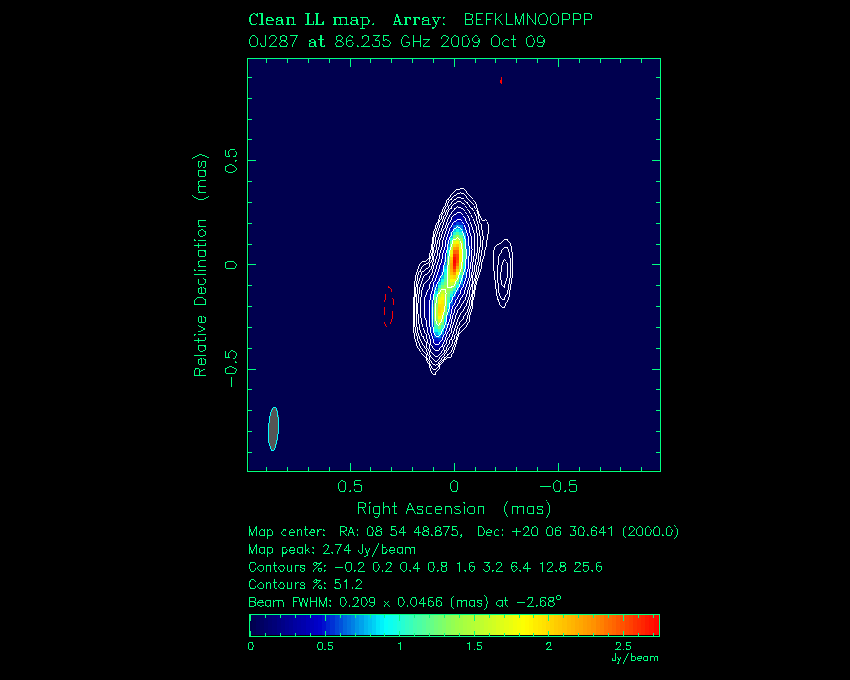}
\caption{Quasi-contemporanous VLBI maps of OJ287 at 15\,GHz (left; data: Mojave data base), 43\,GHz (center, data: Boston group), and 86\,GHz (right, data: GMVA) of October 2009.}
\label{3maps}
\end{figure}
\begin{figure}
\begin{overpic}[trim= 70mm 0mm 50mm 0mm, clip, width=40mm]{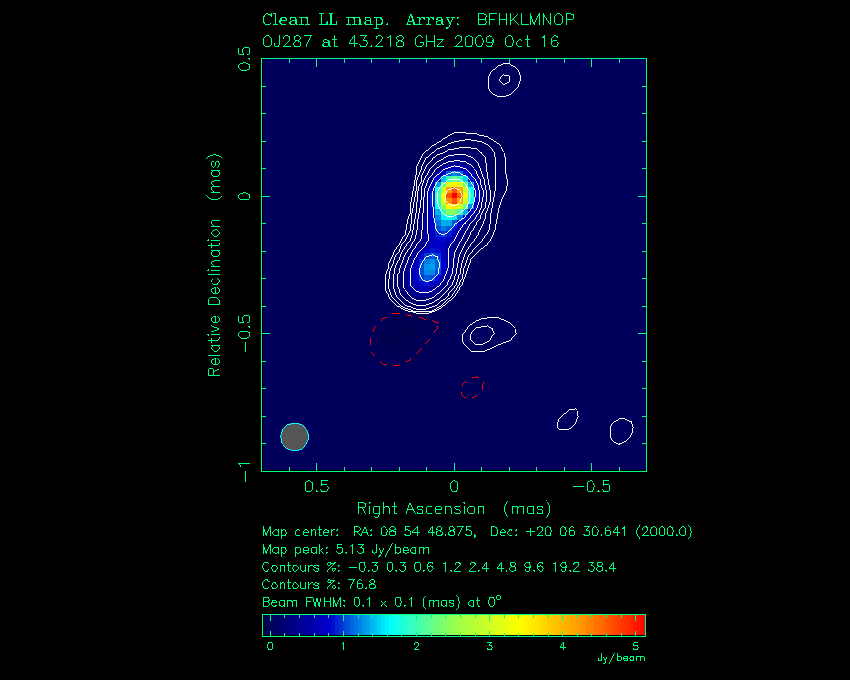}
\put(50,69){\textcolor{red} {\large C1}}
\put(45,56){\textcolor{green} {\large C2}}
\end{overpic}
\begin{overpic}[trim= 70mm 0mm 50mm 0mm, clip, width=40mm]{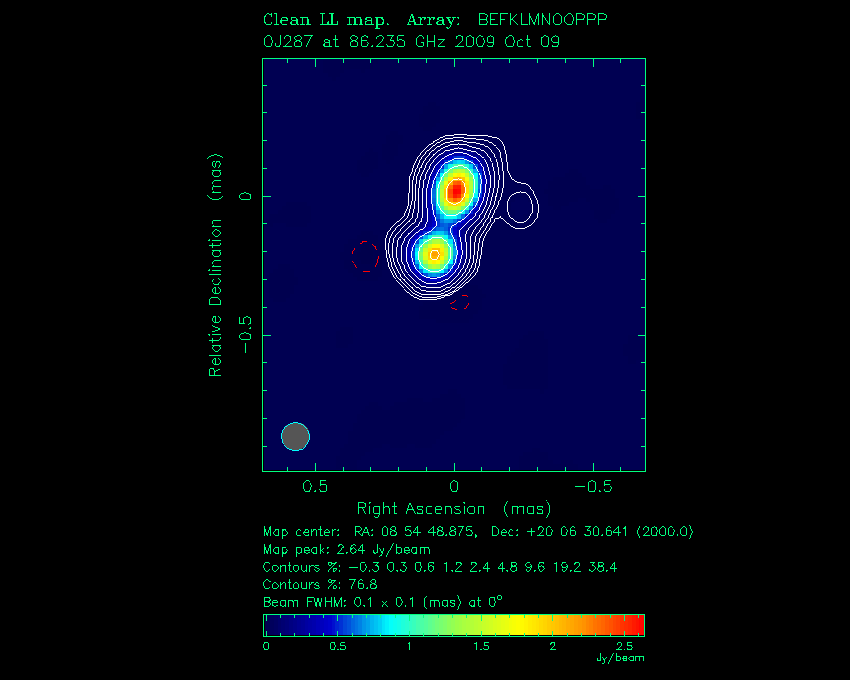}
\put(50,69){\textcolor{red} {\large C1}}
\put(45,56){\textcolor{green} {\large C2}}
\end{overpic}
\includegraphics[width=0.5\textwidth]{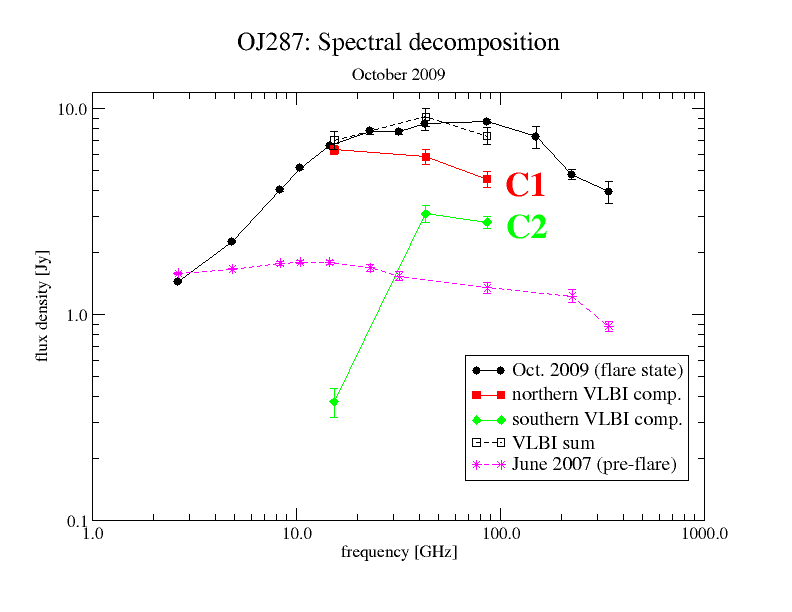}
\caption{Left: 43 GHz VLBI image of OJ287 from VLBA. Center: 86 GHz VLBI image of OJ287 from GMVA. 
Both maps are convolved with a circular beam of 0.1\,mas size for better comparison.
Right: Spectral decomposition. Spectra of the VLBI components C1 and  C2 and the total
radio total spectrum superimposed (black: 2009 - flare state, pink: 2007 - preflare, black dashed: sum of VLBI flux).}
\label{2maps}
\end{figure}
\begin{figure}
\centering\includegraphics[trim= 0mm 0mm 0mm 32mm, clip, width=65mm]{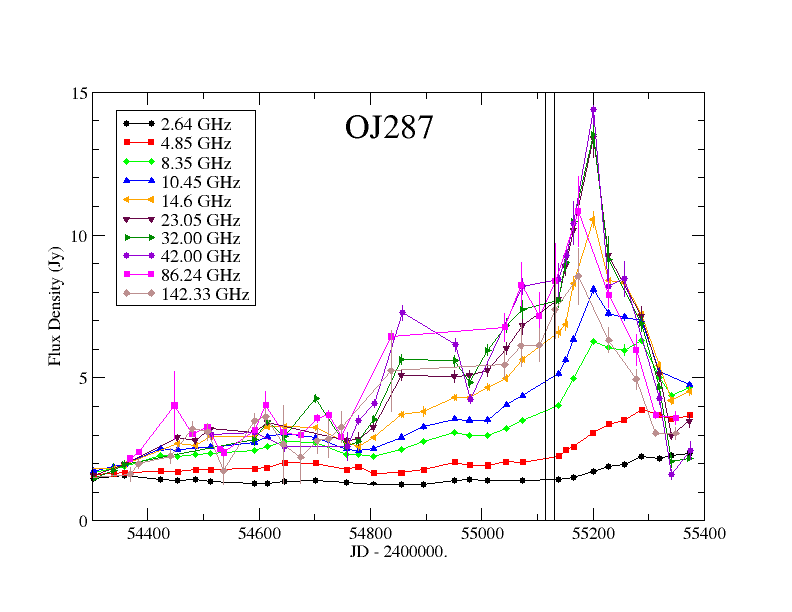}
\vspace* {-5mm}
\caption{Radio variability of OJ287 at cm-wavelengths (data: F-GAMMA monitoring program \cite{fgamma}). A strong 
flux density peak is seen in early January 2010. The two black vertical lines 
bracket the times of the VLBI observations shown in Figure 1.}
\label{oj287flux}
\end{figure}

VLBI imaging with the highest possible angular (and spatial) resolution can shed light on the physical
processes acting within the centers of blazars, in regions where jets are forming and gamma-rays are produced. 
As an example we show in Figure \ref{3maps} three VLBI maps of the prominent blazar OJ\,287 observed in October 
2009 at 15 \& 43\, GHz with the VLBA, and at 86\, GHz with the GMVA. At the lower frequencies, OJ\,287
shows a bent core-jet structure, with a prominent and unresolved core and some
faint jet-emission extending west to $r\leq 5$\,mas. At 86\,GHz, however, the core regions breaks up into
two bright sub-components (C1 \& C2), which interestingly are oriented along a line almost perpendicular to
the direction of the mas-scale jet. If interpreted by a spatially bent jet, one would expect
that the northern most component (C1) to be the VLBI core, which 
is the unresolved base of a synchrotron self-absorbed jet.
In Figure \ref{2maps} (left) we show the 43\,GHz and 86\,GHz maps convolved with a circular
beam of 0.1\,mas in size (factor $\sim 2$ super-resolution at 43\,GHz). The two components
seen at 86\,GHz, now are also clearly visible at 43\,GHz. In October 2009, OJ\,287 was in the rising
phase of a prominent radio-flare (peak: Jan. 2010), which followed a strong
gamma-ray flare (peak: October 24, 2009, see \cite{Agudo}). In Figure \ref{oj287flux} we show the
radio light-curves from the F-GAMMA program \cite{fgamma}, from which we determine two radio spectra, 
one near the onset of the flare and one for October 2009 (see Figure \ref{2maps}, right panel). 
To this figure we have added the 
spectra of the VLBI components C1 and C2, with their flux densities determined from Gaussian
model fits. The spectrum of the fainter southern VLBI component C2 is more inverted and peaks at
a much higher frequency than C1, suggesting that C2 could be the VLBI core and not C1. In view of
the orientation of the pc-scale jet, this however would require some extreme geometry. An alternative
interpretation could be that C2 is a very compact and highly magnetized shock which is evolving.
The continued mm-VLBI monitoring should clarify, if we see a moving or a stationary feature
and if there is a relation with the gamma-ray activity observed at this date.

\section{Global VLBI at 1\,mm wavelength}
\vspace*{-0.4cm}
The development of 1\,mm VLBI started in the the early 1990's and first transatlantic fringes
were detected in 2002/2003 on a 8400\,km long baseline between the IRAM 30\,m telescope (Pico Veleta, Spain)
and the 10\,m SMT (Mt. Graham, AZ, USA) (see: \cite{Krichbaum2008} and references therein).
Recent progress in the development of digital VLBI recording systems has led to
larger observing bandwidth (\& data rates) and higher sensitivities. The VLBI detection
of ultra-compact emission regions in Sgr\,A* and M\,87 at 230\,GHz and with 
telescopes in the USA (SMTO, SMA, CARMA) is another important step towards global mm-/sub-mm
VLBI, which aims at the $\sim 10$\,$\mu$as scale imaging of the event horizon in
nearby black holes (\cite{Fish}, Event Horizon Telescope:
\href{http://eventhorizontelescope.org}{EHT}).

\begin{figure}
\begin{minipage}[t!]{0.99\textwidth}{
\includegraphics[trim= 0mm 90mm 0mm 0mm, clip, width=0.32\textwidth]{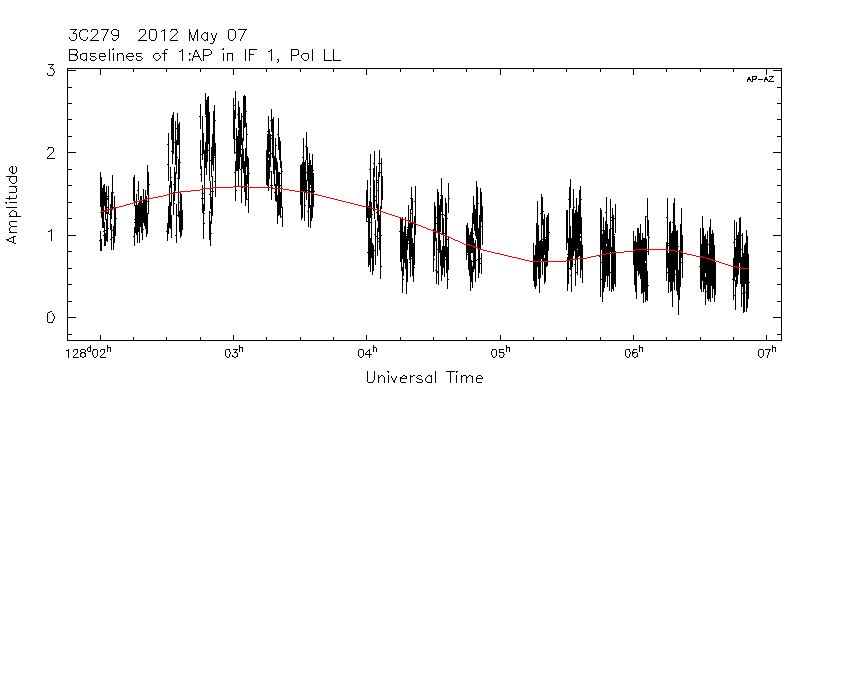}
\includegraphics[trim= 0mm 90mm 0mm 0mm, clip, width=0.32\textwidth]{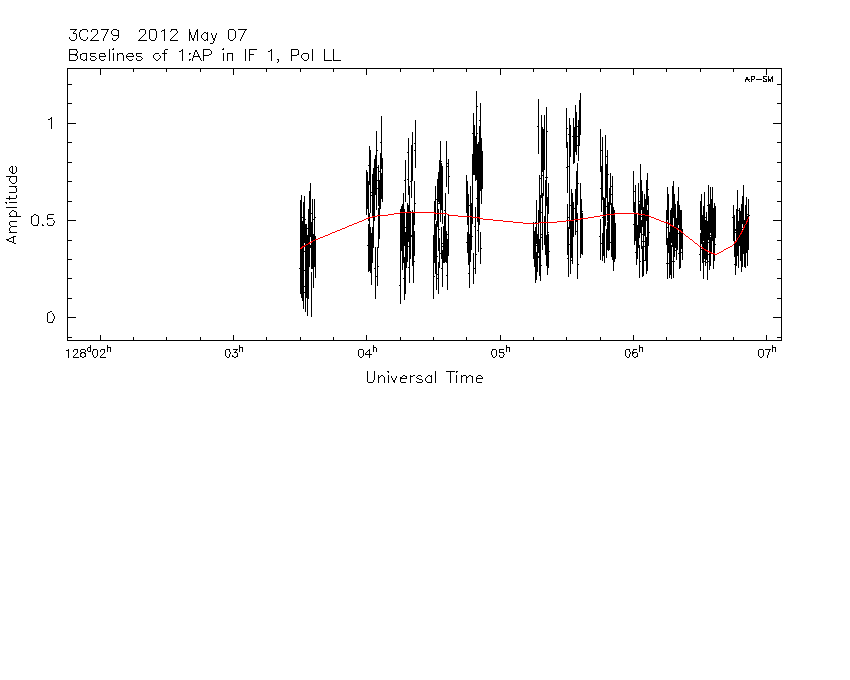}
\includegraphics[trim= 0mm 90mm 0mm 0mm, clip, width=0.32\textwidth]{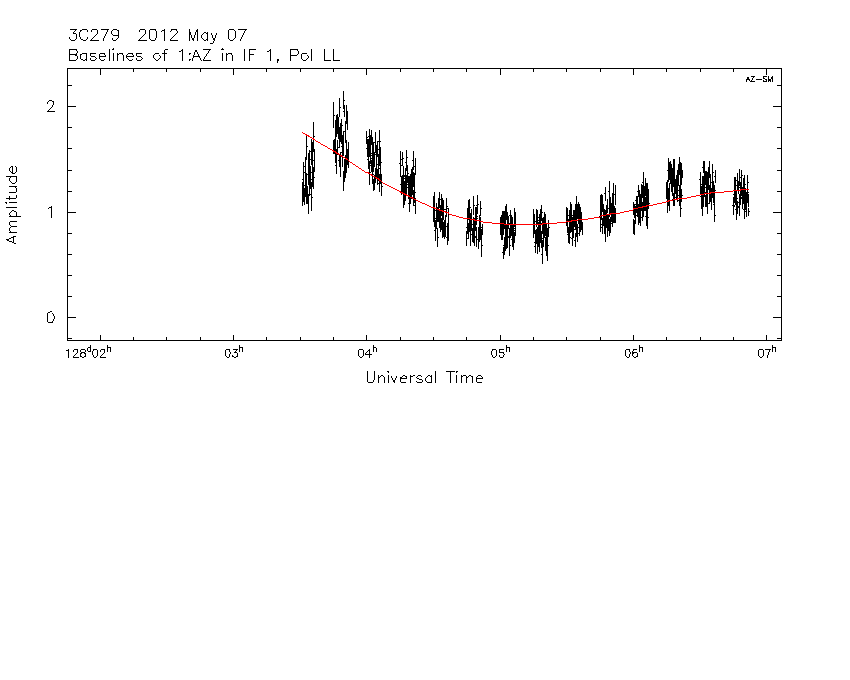}
}
\end{minipage}
\begin{minipage}{0.99\textwidth}{
~~\\
\hspace*{0mm}\includegraphics[width=0.33\textwidth]{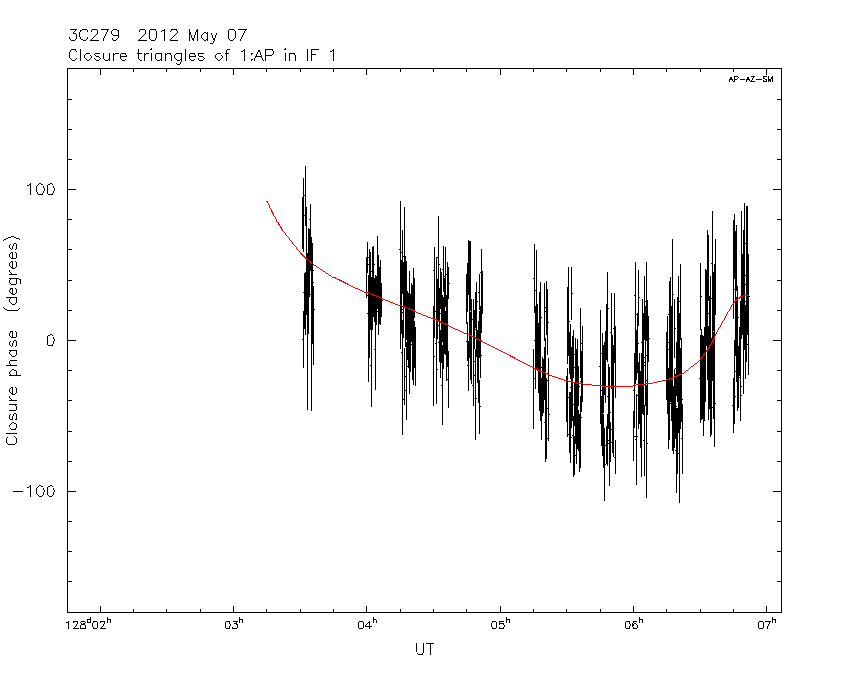}
\includegraphics[width=0.33\textwidth]{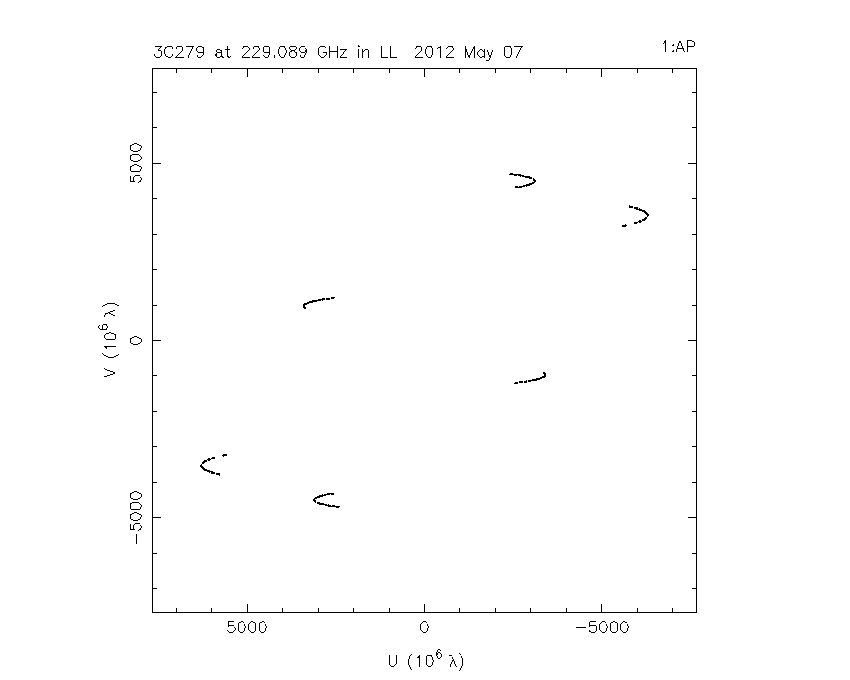}
}
\end{minipage}
\caption{Top: Visibility amplitudes versus time at 230\,GHz and
for the baselines APEX-SMTO (left), APEX-SMA (center),
and SMTO-SMA (right). Bottom, left: Closure phase for this triangle. The red line shows the fit 
of the Gaussian model displayed in Fig. 5. Bottom, center: actual uv-coverage.}
\label{1mmvis}
\end{figure}

\begin{figure}
\includegraphics[trim= 65mm 0mm 50mm 11mm, clip, width=0.3\textwidth]{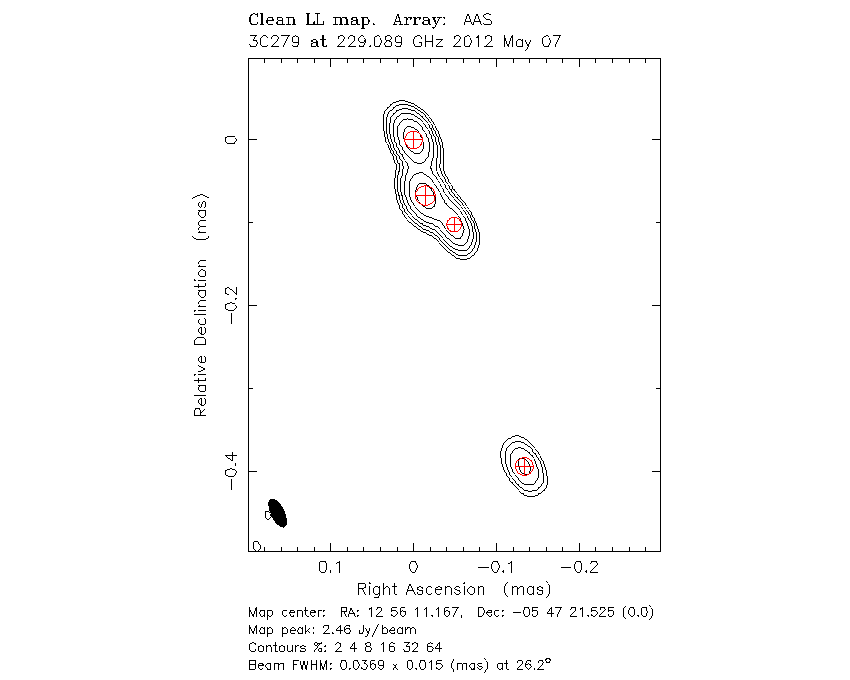}
\hspace*{20mm}\includegraphics[width=0.5\textwidth]{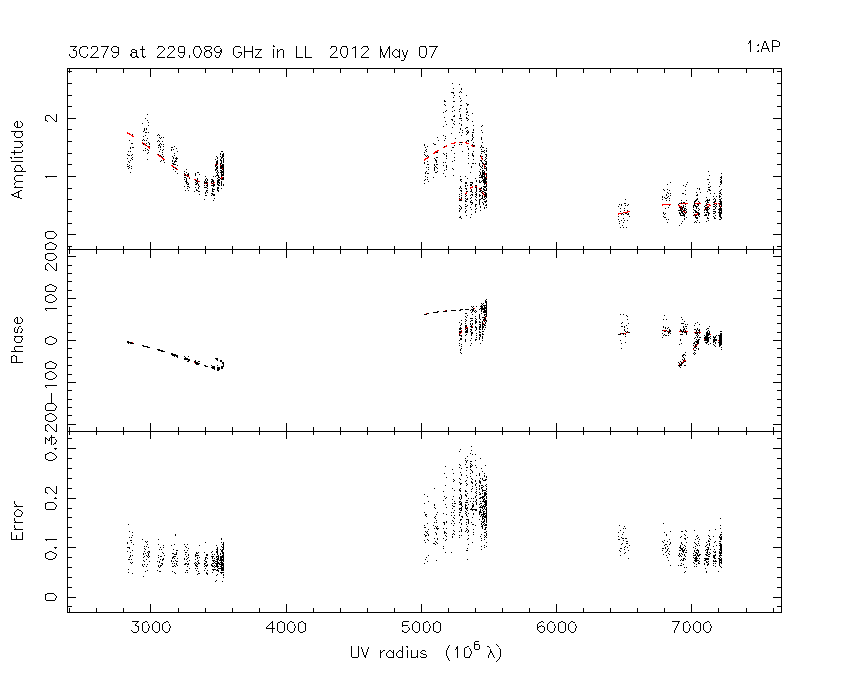}
\caption{Left: Map of Gaussian modelfit components for 3C279 at 230 GHz. Four circular 
Gaussians, indicated by red circles with crosses, are fitted to the visibilities, see Fig. 4.
The map is convolved with a uniform weighted observing beam (FWHM: 37 x 15 $\mu$as).
Right: Visibility amplitude and phase plotted versus uv-distance. The red dashed lines 
show the fit of the Gaussian model to the data.}
\label{1mmmap}
\end{figure}

\begin{figure}
\centering\includegraphics[width=0.3\textwidth]{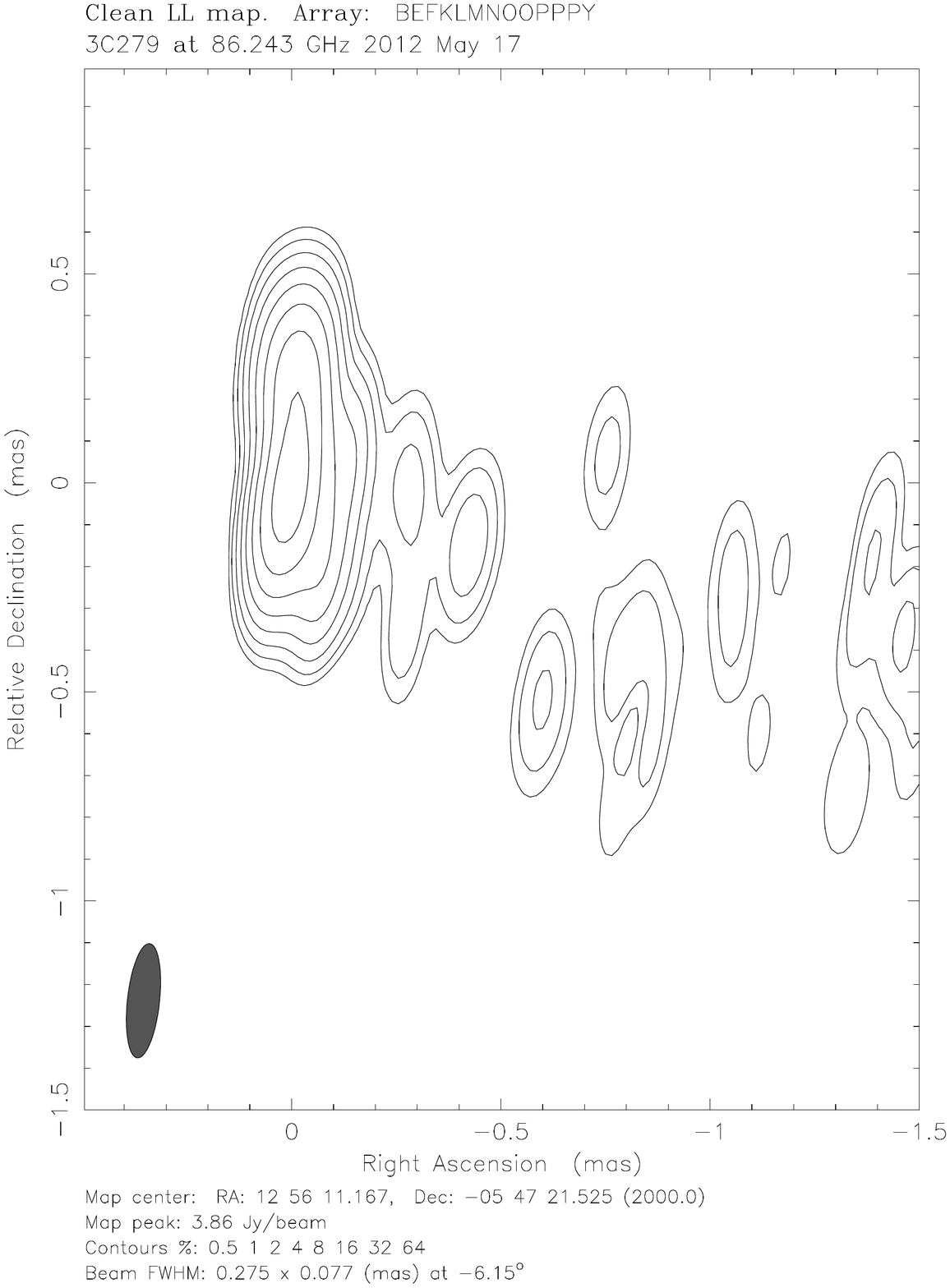}
\centering\includegraphics[width=0.55\textwidth]{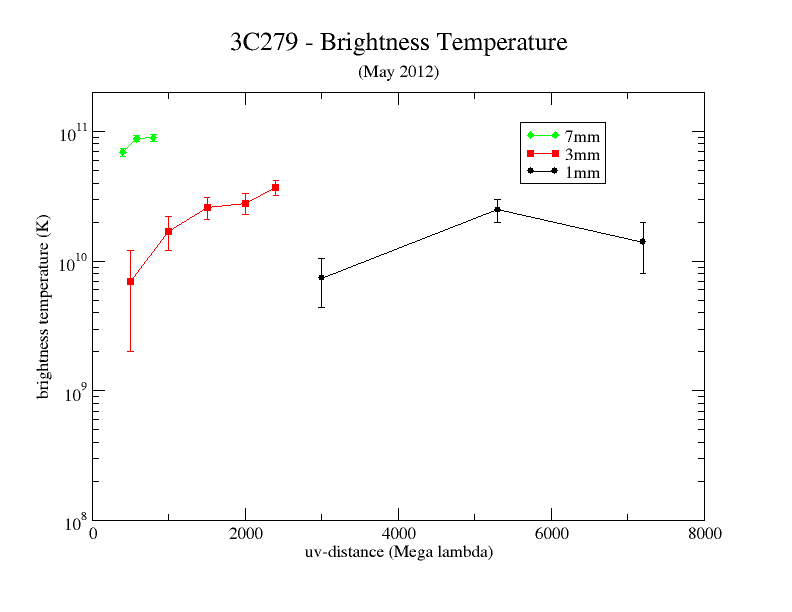}
\caption{Left: 3mm VLBI map of 3C\,279 observed with the GMVA on May 17, 2012. The
size of the uniform weighted clean beam is (275 x 77) $\mu$as. Right: Brightness temperature
of 3C\,279 at 43\,GHz (green), 86\,GHz (red) and 229\,GHz (black) plotted versus projected
baseline length (in M$\lambda$). The data are all quasi-simultaneously obtained in early May 2012.}
\label{1mmtb}
\end{figure}

In this context we begun in 2009 the planning, and in 2010 the outfit 
of the 12\,m APEX telescope in Chile (Llano de Chajnantor,
5105\,m) for VLBI. In addition to the science case (see the Whitepaper \cite{Whitepaper})
our second motivation was the idea that APEX can serve as pathfinder
for future mm-VLBI with ALMA. It took 2 years of work and one 
unsuccessful VLBI-test (in spring 2011), before we could claim 
the detection of first fringes with the APEX telescope. The setup and the observing
details are presented in the talk by A. Roy et al. (this conference \cite{Roy}), and \cite{Wagner}).
On May 7, 2012 the bright quasar 3C\,279 (S$_{\rm 230\,GHz}$ = 19.8\,Jy)
was observed with APEX and the SMTO (Arizona) in circular polarization and
with the phased SMA (Hawaii) in linear polarization at a 
recording rate of 2 GBit/s. On the baselines to APEX, fringes were detected with
a typical SNR $\sim 10-15$ and on the SMTO-SMA baseline the source was seen with 
SNR $\sim 30-35$. The detection of 3C\,279 on the 7.22G$\lambda$ long VLBI baseline between
Chile and Hawaii forms a new world-record in angular resolution (28.6\,$\mu$as)
and demonstrates the feasibility of 1.3\,mm VLBI even on the longest baselines.
The amplitude calibration of the visibilities took into account all known effects
and resulted in the visibilities shown in Fig. \ref{1mmvis} (see \cite{Wagner} for details).
In Fig. \ref{1mmmap} we show a preliminary 1\,mm VLBI map of 4 circular Gaussian 
components, which fit the visibilities and closure phases reasonably well\footnote{Linear 
polarization at SMA may affect the closure phase if 3C\,279 is highly polarized or antenna D-terms are large.}  
($\chi^2_{\rm reduced}=2.42$). 
Owing to the sparse uv-coverage and 
the residual calibration uncertainty, this model is not unambiguous and should be regarded with some care. 
A robust statement however seems to be, that 3C\,279 appears elongated in North-South direction on 
the 0.1\,mas scale. A similar elongation is also seen in a 3\,mm VLBI image obtained 
with the GMVA only 10\,days later (Fig. \ref{1mmtb}, left panel). In the imaging process we 
were puzzled about a relatively low correlated flux at 230\,GHz 
of $\sim 1-2$\,Jy at $\sim 3$\,G$\lambda$ and  $\sim 0.7 - 1.2$\,Jy at $\sim 7$\,G$\lambda$ 
(see Fig. \ref{1mmmap}, right panel). In Fig.\ref{1mmtb} (right panel)
we therefore plot the brightness temperature for a few typical uv-distances. We also have added 
near-in time VLBI data at 86\,GHz (GMVA) and 43\,GHz (VLBA). It can be seen that the
brightness temperatures at 86\,GHz and 230\,GHz are very similar, but about a factor of
$5-10$ lower than at 43\,GHz. A jet brightness temperature which is decreasing with increasing frequency
could indicate intrinsic jet acceleration or a decrease of the jet Doppler-factor towards
the black hole.

\section{Future Perspective}
\vspace*{-0.4cm}
The obvious next steps for improving the capabilities of mm-VLBI are the addition
of more sensitive telescopes and the further enhancement of array sensitivities.
Since at mm-wavelengths weather is a limiting factor, triggered VLBI observations
in pre-allocated longer time blocks are very feasible and require only moderate
logistical efforts at the individual stations.
The addition of the phased ALMA, equi\-va\-lent to a $\sim 85$\,m dish, to the existing VLBI arrays 
(at $\nu \geq 43$\,GHz) will lower the detection limits for mm-VLBI by up to 2 orders of magnitude 
(see \cite{Alef} for the ALMA Phasing Project (APP)). Global mm-VLBI with ALMA however will be only possible,
if in parallel to the ongoing technical development also a strong scientific user community not only obtains interesting results using the existing mm-VLBI arrays, but also actively pushes ALMA to this new frontier.

\end{document}